\documentclass[conference]{IEEEtran}

\usepackage{multirow}     %
\usepackage{booktabs}     %
\usepackage{amsmath,amssymb,amsfonts}
\usepackage{rotating}
\usepackage{graphicx}
\usepackage{algorithm}
\usepackage{algorithmic}
\usepackage{adjustbox}
\usepackage[graphicx]{realboxes}
\usepackage{subfigure}
\usepackage{cite}
\usepackage{float}
\usepackage{textcomp}
\usepackage{xcolor}
\usepackage{colortbl}

\def\BibTeX{{\rm B\kern-.05em{\sc i\kern-.025em b}\kern-.08em
    T\kern-.1667em\lower.7ex\hbox{E}\kern-.125emX}}
\begin{document}

\bstctlcite{IEEEexample:BSTcontrol}
\title{CIMR-V: An End-to-End SRAM-based CIM Accelerator with RISC-V for AI Edge Device\\
}
\author{\IEEEauthorblockN{Yan-Cheng Guo\textsuperscript{1} and, Tian-Sheuan Chang\textsuperscript{1}, \textit{Senior Member, IEEE},  \\
Chih-Sheng Lin\textsuperscript{2}, Bo-Cheng Chiou\textsuperscript{2}, Chih-Ming Lai\textsuperscript{2}, Shyh-Shyuan Sheu\textsuperscript{2}, 
 Wei-Chung Lo\textsuperscript{2} and Shih-Chieh Chang\textsuperscript{2}
 }
\IEEEauthorblockA{\textit{\textsuperscript{1}{Institute of Electronics, National Yang Ming Chiao Tung University,}}
Hsinchu, Taiwan
}
\IEEEauthorblockA{\textit{\textsuperscript{2}{Industrial Technology Research Institute, }}
Hsinchu, Taiwan
}
}
\maketitle

\begin{abstract}%

Computing-in-memory (CIM) is renowned in deep learning due to its high energy efficiency resulting from highly parallel computing with minimal data movement. 
However, current SRAM-based CIM designs suffer from long latency for loading weight or feature maps from DRAM for large AI models.
Moreover, previous SRAM-based CIM architectures lack end-to-end model inference.
To address these issues, this paper proposes CIMR-V, an end-to-end CIM accelerator with RISC-V that incorporates CIM layer fusion, convolution/max pooling pipeline, and weight fusion, resulting in an 85.14\% reduction in latency for the keyword spotting model.
Furthermore, the proposed CIM-type instructions facilitate end-to-end AI model inference and full stack flow, effectively synergizing the high energy efficiency of CIM and the high programmability of RISC-V.
Implemented using TSMC 28nm technology, the proposed design achieves an energy efficiency of 3707.84 TOPS/W and 26.21 TOPS at 50 MHz.

~\\
\noindent Keywords : Computing-in-memory, AI accelerator, Pruning framework
\end{abstract}

\section{Introduction}

In recent years, convolution neural networks (CNNs) have had a profound impact on various applications, including but not limited to image classification~\cite{2017_ImageNet} and speech processing\cite{2014_KWS}, owing to their remarkable ability to extract features through very deep layers.
However, the use of very deep layers in CNNs increases the memory bandwidth and requires larger computation demands, leading to massive data movement between memory and processing elements (PEs). This results in high latency and power consumption in current digital deep learning accelerators (DLAs).
CIM offers an alternative approach by combining storage and computation, resulting in highly parallel computing and high energy efficiency~\cite{2021_CIM_summary}.

\begin{figure}[t]
\centering
\includegraphics[width=1.0\linewidth]{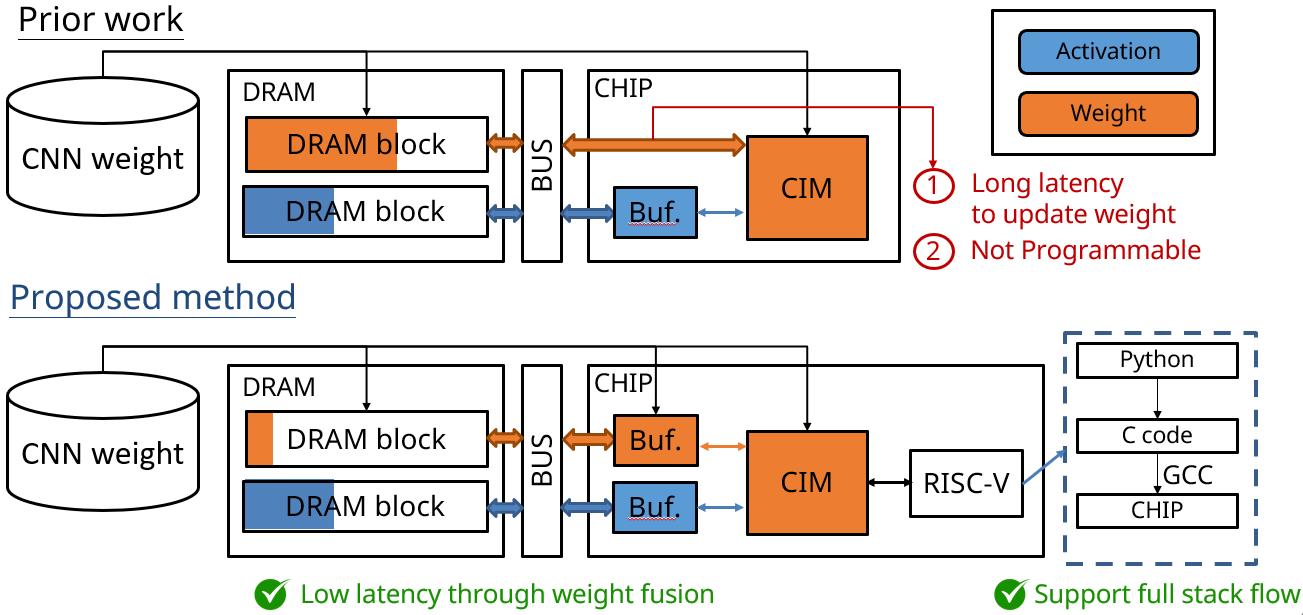}
\caption{The previous work failed to address the long latency involved in weight updates and lacked programmability. In contrast, our method incorporates a weight buffer to reduce latency and supports a full-stack flow integrating the RISC-V and CIM architecture.}
\label{00_motivation}
\end{figure}

However, previous SRAM-based CIM designs \cite{2021_JSSC_SRAM,2022_TCAS_SRAM} primarily focus on reducing the power consumed during weight data movement, while failing to address the need to minimize feature map data movement throughout the entire model execution. (Fig.~\ref{00_motivation})
Moreover, as these deep learning models become deeper and larger to achieve higher accuracy, existing CIM designs with small array sizes will require frequent updates of model weights from DRAM, leading to high latency.

We propose CIMR-V to overcome the limitations mentioned above by reducing feature map data movement through CIM layer fusion and pipeline max pooling block, and reducing weight loading latency using the weight fusion method. 
Additionally, CIMR-V allows for user-friendly AI model deployment via the full stack flow.
Implemented in TSMC 28nm technology, the proposed design achieves superior performance, with an energy efficiency of 3707.84 TOPS/W and a throughput of 26.21 TOPS at 50 MHz.

The remainder of the paper is organized as follows. Section II shows the proposed hardware architecture and data flow. Section III presents the experimental results. Finally, this paper is concluded in Section IV.

\section{Proposed Hardware Architecture and Data Flow}

\subsection{The Overall Architecture of CIMR-V.}
\begin{figure}[htbp]
\centering
\includegraphics[height=!,width=0.8\linewidth]{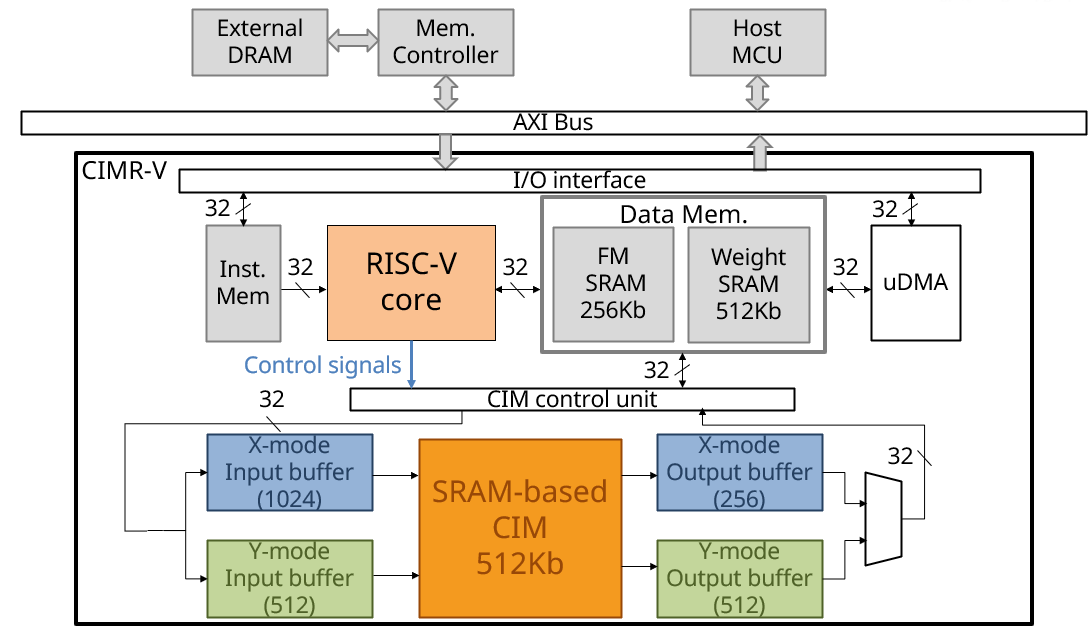}
\caption{The overall architecture of CIMR-V.}
\label{02_architecture}
\end{figure}

Fig.~\ref{02_architecture} depicts the comprehensive architecture of CIMR-V, which is composed of a 512Kb SRAM-based CIM unit for energy-efficient computing, a modified RISC-V core for controlling the CIM and conducting high-precision computing, an instruction memory, a 256Kb feature map SRAM for layer fusion, and a 512Kb weight SRAM for weight fusion. 
CIMR-V is implemented on the PULPissimo platform, featuring the ibex 32-bit 2-stage RISC-V architecture \cite{2018_PULPissimo_RISCV}.
In order to reduce routing complexity and power consumption, the CIM input buffer is designed with a 32-bit shift.

\subsection{The High-density SRAM-based CIM Macro with X-mode and Y-mode}

This design is based on the SRAM-based CIM~\cite{2021_ITRI_CIM_SRAM}, which supports two distinct modes:
\begin{enumerate}
\item X-mode, which accommodates high input data with 1024 wordlines (WLs), 512 bitlines (BLs), and 256 sense amplifiers (SAs).
\item Y-mode, which accommodates high output data with 512 WLs, 1024 BLs, and 512 SAs.
\end{enumerate}

To minimize power consumption and quantization error, the output of 512 / 1024 MAC operations is sensed by the SA on the corresponding long BL, and the activation function (ReLU) is executed simultaneously. 
Additionally, to enhance the robustness and accuracy of CNNs, we apply the symmetry weight mapping method to mitigate nonlinearity (NL) and cell variation in binary or ternary weights.

\subsection{The Modified RISC-V Core with CIM Instruction}
\begin{figure}[t]
\centering
\includegraphics[width=1.0\linewidth]{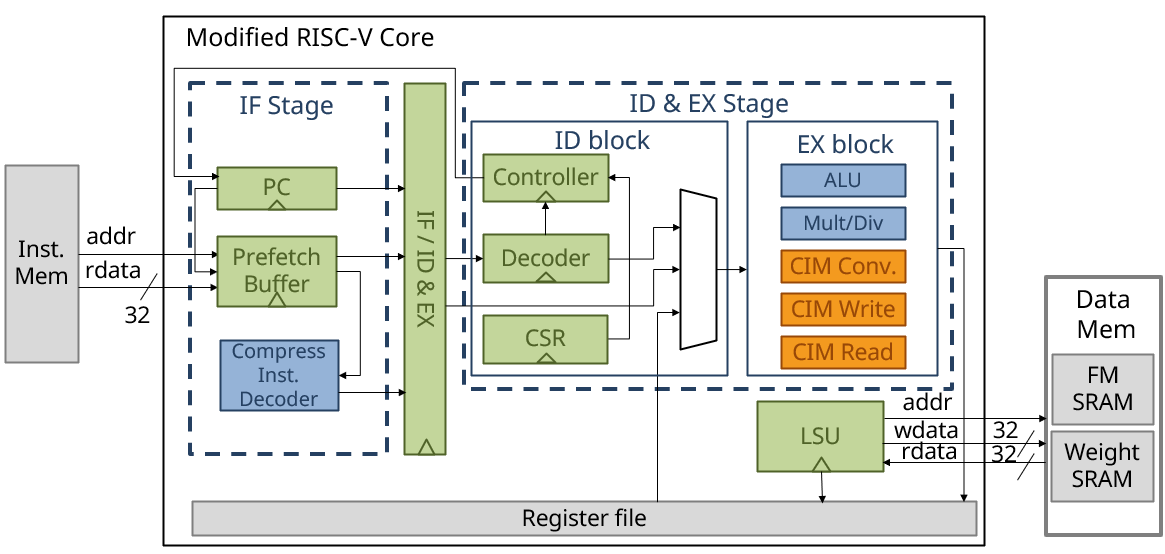}
\caption{The modified RISC-V core with CIM instruction for AI model inference.}
\label{02_Modified_RISCV_core}
\end{figure}
Fig.~\ref{02_Modified_RISCV_core} illustrates the modified RISC-V core, which incorporates not only the native Arithmetic Logic Unit (ALU) for basic arithmetic operations but also energy-efficient components, specifically the CIM read/write and CIM convolution modules.

To reduce pipeline overhead, we employ a two-stage CPU architecture. 
In the first stage, an instruction is fetched from memory using a prefetch buffer. 
In the second stage, the fetched instruction is decoded by the decoder unit, and the controller adjusts the program counter (PC) address and the control and status register (CSR) as needed. 
Simultaneously, the Load Store Unit (LSU) unit simultaneously loads data from data memory into the register file. 
Finally, in the execute block, the input data is computed and the output is stored in the register file.

To achieve optimal efficiency, all CIM instructions, such as the CIM convolution, write, and read instructions, are executed atomically within a single cycle, thereby minimizing latency and storage overhead. 
Additionally, to reduce feature map data movement, the CIM instructions utilize data from the feature map SRAM or weight SRAM instead of the register file.

\subsection{The Energy-efficient CIM-type Instruction}
\begin{figure}[t]
\centering
\includegraphics[width=1.0\linewidth]{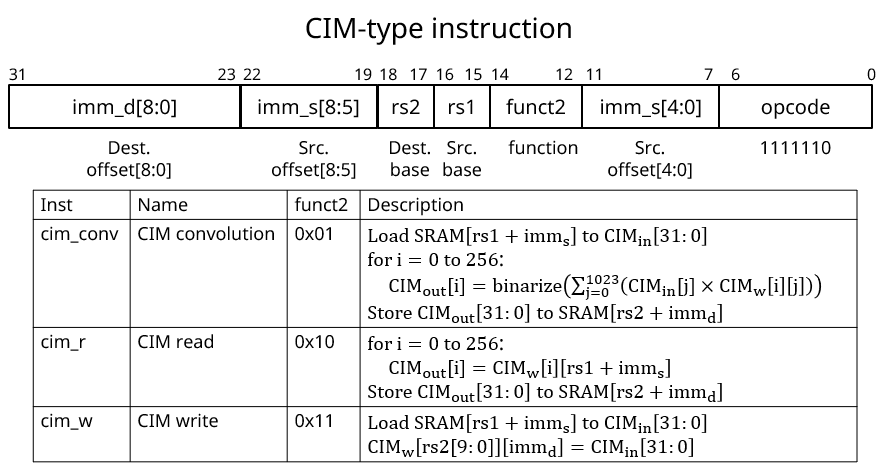}
\caption{The energy-efficient CIM-type instruction for CIMR-V.}
\label{04_cim_inst}
\end{figure}

Fig.~\ref{04_cim_inst} illustrates the energy-efficient CIM-type instruction (opcode 1111110) employed in CIMR-V, which includes three main instructions: CIM convolution, CIM read, and CIM write. 
To improve system throughput, the CIM convolution instruction is executed within a single cycle by loading FM SRAM data from address register 1 (rs1) with the immediate source offset ($imm_{s}$) into the CIM input buffer, performing 512/1024 Multiply-Accumulate (MAC) operations, and storing the resulting CIM output into FM SRAM at rs2 with the added offset $imm_{d}$. 

Similarly, the CIM read instruction reads the CIM macro weights from rs1 + $imm_{s}$ and stores them in SRAM at rs2 + $imm_{d}$. 
Additionally, the CIM write instruction supports the writing of 32-bit data in a single operation, further enhancing writing performance. 
Overall, this instruction set contributes to the energy efficiency and high performance of CIMR-V.

\subsection{Row-wise Convolution Dataflow: CIM Layer Fusion and Pipeline with Convolution and Max Pooling}
\begin{figure}[t]
\centering
\includegraphics[width=1.0\linewidth]{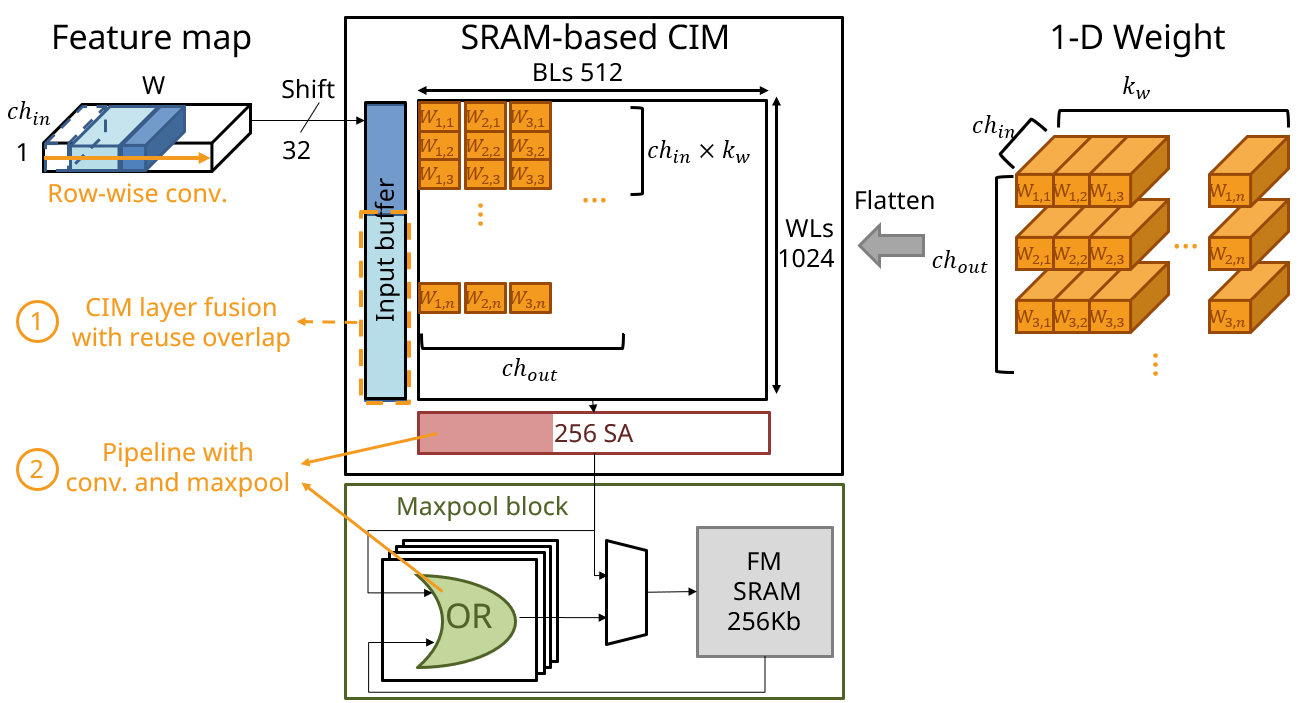}
\caption{The row-wise convolution dataflow enables CIM layer fusion\cite{2016_layer_fusion_for_digital_AI_accelerator} and pipeline with convolution and max pooling.}
\label{05_dataflow_1D_conv}
\end{figure}

\begin{figure}[t]
\centering
\includegraphics[width=1.0\linewidth]{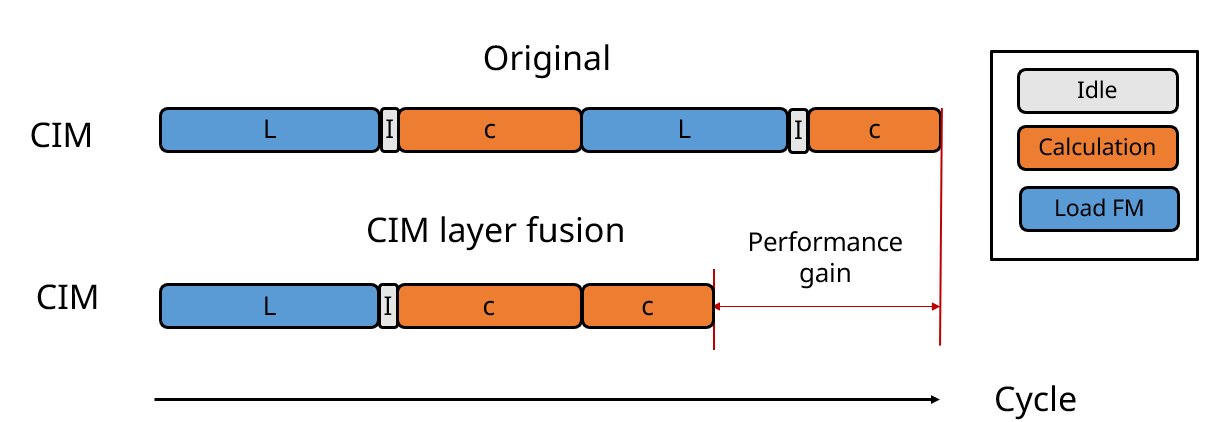}
\caption{The performance gain example of CIM layer fusion.}
\label{CIMR-V_08_CIM_layer_fusion_schedule}
\end{figure}

\begin{figure}[t]
\centering
\includegraphics[width=1.0\linewidth]{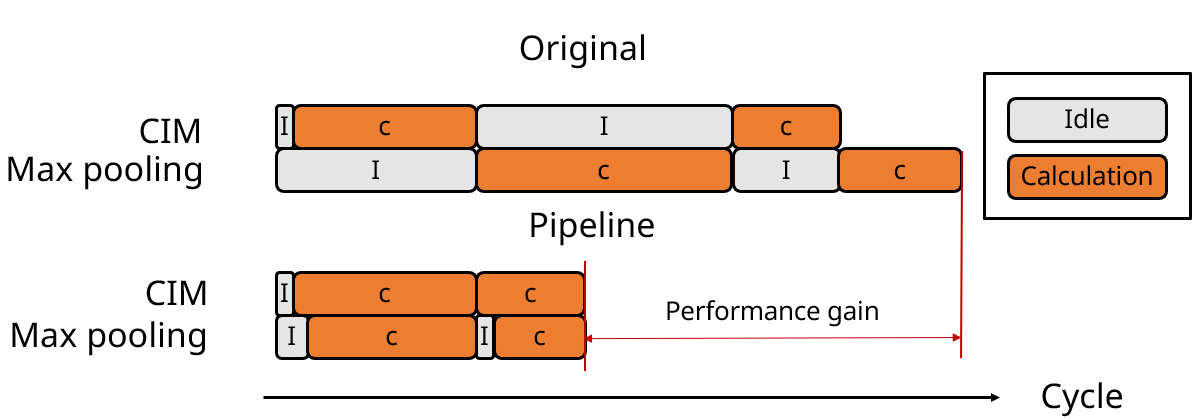}
\caption{The performance gain example of convolution/max-pooling pipeline.}
\label{CIMR-V_08_pipeline_schedule}
\end{figure}

The overall dataflow for the CIM convolution instruction is depicted in Fig.~\ref{05_dataflow_1D_conv}. This involves shifting the input feature map into the input buffer, flattening the CNN weights into macro BLs by output channel, and executing the convolution to obtain output data through SA.

To minimize power consumption and feature map data movement, the CIM convolution instruction supports CIM layer fusion, achieved by dividing the input feature map into blocks and storing overlapping data through the CIM input buffer to reduce layer fusion overhead. (Fig.~\ref{CIMR-V_08_CIM_layer_fusion_schedule})
Furthermore, CIMR-V supports a pipeline with CIM convolution and max pooling to reduce latency and storage space. (Fig.~\ref{CIMR-V_08_pipeline_schedule})

In general, row-wise convolution dataflow is crucial for enhancing throughput by utilizing CIM layer fusion and convolution/maxpool pipeline technology.
Moreover, the dataflow is highly scalable to row-wise 2-D convolution.

\subsection{Weight Fusion to Minimize Weight Loading Latency}

\begin{figure}[t]
\centering
\includegraphics[width=0.5\linewidth]{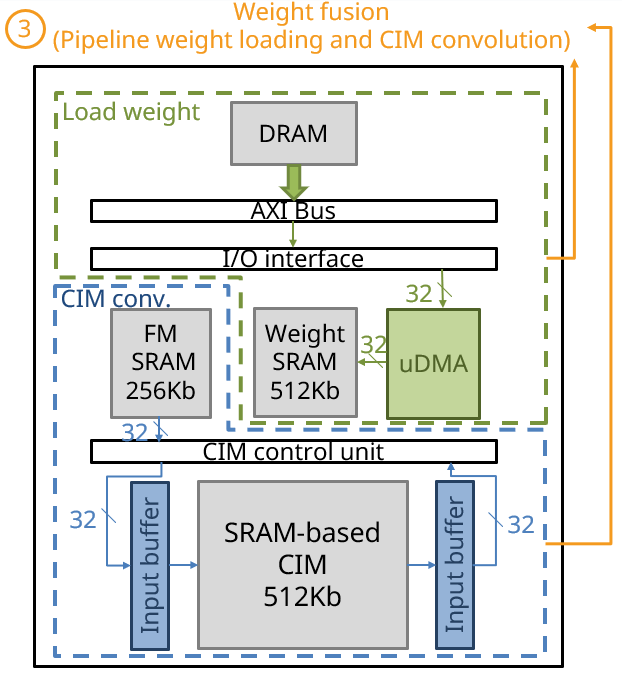}
\caption{The weight fusion pipeline combines weight loading and CIM convolution to minimize latency caused by weight loading from DRAM.}
\label{06_weight_fusion}
\end{figure}

\begin{figure}[t]
\centering
\includegraphics[width=1.0\linewidth]{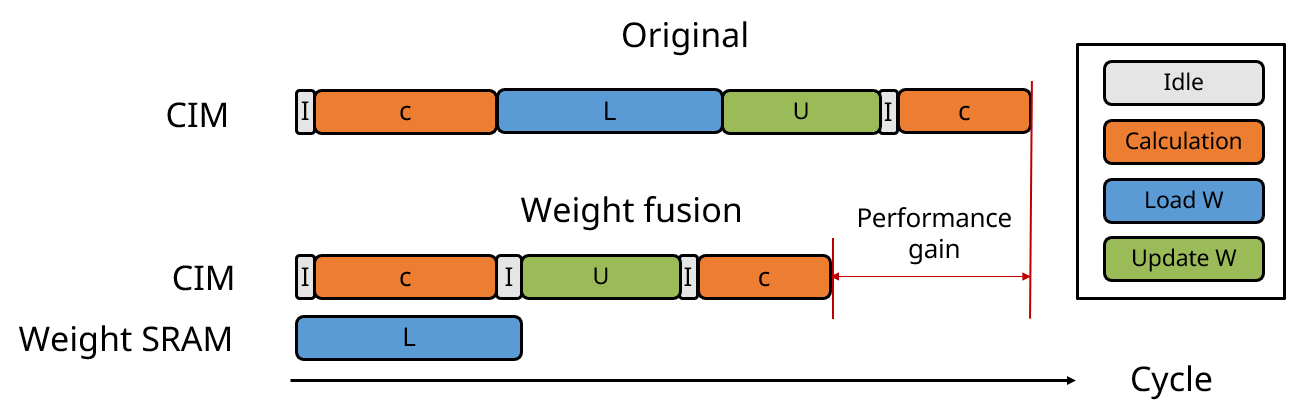}
\caption{The performance gain example of weight fusion.}
\label{CIMR-V_08_weight_fusion_schedule}
\end{figure}

Fig.~\ref{06_weight_fusion} depicts the proposed weight fusion method for CIM, which significantly reduces the weight loading latency from DRAM. 
This is achieved by pipelining the CIM convolution and weight loading into the weight SRAM, resulting in faster weight data access. 

Furthermore, to minimize CPU loading, we employ uDAM for the parallel loading of weight data.
uDAM is a technique that allows for direct data transfer between devices without the intervention of the CPU. 
This approach enables efficient weight data transfer and reduces CPU overhead. 

Finally, the updated weight data from the weight SRAM is written into the CIM macro using the CIM write instruction. 
The proposed method offers a more efficient and streamlined weight loading process for CIM convolution, enhancing the overall performance of the system. (Fig.~\ref{CIMR-V_08_weight_fusion_schedule})

\subsection{Full Stack Flow with CIMR-V for AI Models}

CIMR-V offers a full stack flow for model deployment, distinguishing it from conventional AI accelerators. 
The workflow involves converting the model's Python code, trained on TensorFlow or PyTorch, to C/C++ for deployment on CIMR-V using the GCC compiler and RISC-V architecture. 

Furthermore, the full stack flow allows for seamless and efficient model deployment on CIMR-V, where CIM-type instructions effectively execute computation. 
This feature enables CIMR-V to deliver high performance with minimal latency, making it an excellent choice for AI workloads.

\subsection{End-to-end AI Model Inference Flow of CIMR-V}
\begin{figure}[htbp]
\centering
\includegraphics[height=!,width=0.7\linewidth]{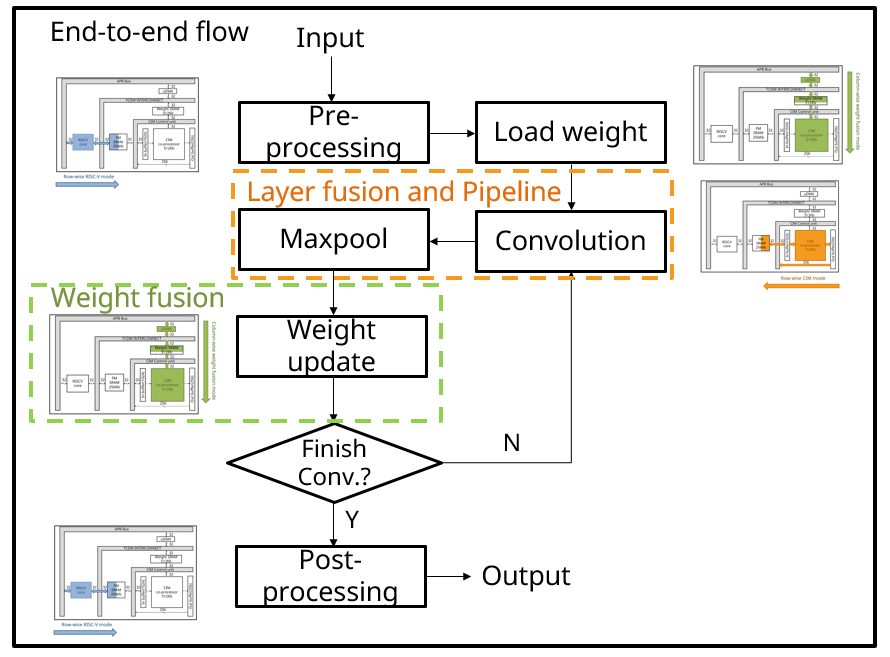}
\caption{CIMR-V end-to-end flow for AI model inference.}
\label{06_end_to_end_flow}
\end{figure}

\begin{table*}[htb]
\caption{Summary of performance and comparison with other designs.}
\label{tb01_HW_comparison}
\centering
\begin{tabular}{|c|c|c|c|c|c|}
\hline
 & JSSC’21\cite{2021_JSSC_SRAM}& TCAS-1’22\cite{2022_TCAS_SRAM}&ISSCC’22\cite{2022_ISSCC_DIANA} & \textbf{This work} \\ \hline
Technology & 65nm & 28nm & 22nm & \textbf{28nm}  \\ \hline
Memory type      
& 6T SRAM              
& 6T SRAM               
& 6T SRAM
& \textbf{10T SRAM}  \\ \hline

\begin{tabular}[c]{@{}c@{}}Array size (WL×BL×sub-bank)\end{tabular} 
& \begin{tabular}[c]{@{}c@{}}128Kb (512×256×1)\end{tabular}      
& \begin{tabular}[c]{@{}c@{}}64Kb (16×64×16)\end{tabular}         
& \begin{tabular}[c]{@{}c@{}}576Kb (1152×512×1)\end{tabular}                  
& \textbf{\begin{tabular}[c]{@{}c@{}} 512Kb (1024×512×1)\end{tabular}} \\ \hline

IA(bits) & 4b/8b & 1b-8b & \begin{tabular}[c]{@{}c@{}}2b/4b/8b(digital) / 7b(analog)\end{tabular}   & \textbf{1b} \\ \hline
W(bits) & 4b/8b & 1b-8b &  \begin{tabular}[c]{@{}c@{}}2b/4b/8b(digital) / 1.5b(analog)\end{tabular}  & \textbf{1b} \\ \hline
OA(bits) & - & 6b & - & \textbf{1b} \\ \hline
Supply voltage (V) & 1 & 0.8 & 0.55 & \textbf{0.9} \\ \hline
Frequency (MHz) & 1000 & 333.33 & 50-320 & \textbf{50} \\ \hline

Throughput (TOPS) 
& \begin{tabular}[c]{@{}c@{}}0.0055\\ (0.352\footnotemark[1])\end{tabular}
& -
& \begin{tabular}[c]{@{}c@{}}29.5 \\ (309.75\footnotemark[1])\end{tabular}
& \textbf{\begin{tabular}[c]{@{}c@{}}26.21\\ (26.21\footnotemark[1])\end{tabular}} \\ \hline

Energy efficiency (TOPS/W)
& \begin{tabular}[c]{@{}c@{}}0.91\\ (166.91\footnotemark[2])\end{tabular}
& \begin{tabular}[c]{@{}c@{}}1280\\ (1011.36\footnotemark[2])\end{tabular}
& \begin{tabular}[c]{@{}c@{}}600\\ (1848.61\footnotemark[2])\end{tabular}
& \textbf{\begin{tabular}[c]{@{}c@{}}3707.84\\ (3707.84\footnotemark[2])\end{tabular}} \\ \hline

Algorithm
& RNN & CNN & CNN & \textbf{CNN}  \\ \hline

Dataset
& GSCD & CIFAR100 & CIFAR10 & \textbf{GSCD}  \\ \hline

Accuracy
& 92.75\% & 76.40\% & 89.3\%-91.4\% & \textbf{94.02\%}  \\ \hline

\begin{tabular}[c]{@{}c@{}}End-to-end \end{tabular}                    
& - & - & \checkmark & \textbf{\checkmark} \\ \hline
\begin{tabular}[c]{@{}c@{}}Weight fusion \end{tabular}                   
& - & - & - & \textbf{\checkmark} \\ \hline

\end{tabular}

\footnotemark[1]{Normalized operations = operations$\times$ activation precision$\times$ weight precision}

\footnotemark[2]{Normalized energy efficiency = energy efficiency$\times$ activation precision$\times$ weight precision$\times \left(\frac{process}{28nm}\right) \times \left(\frac{voltage}{0.9V}\right)^2$}
\end{table*}

Fig.~\ref{06_end_to_end_flow} presents the CIMR-V end-to-end flow for model inference through the RISC-V mode, the CIM mode, and the weight fusion mode. The pre-processing and post-processing steps that need high precision will be executed on RISC-V. The convolutional and pooling layers are executed through the CIM mode but with pipeline to reduce latency. Furthermore, the full model can be executed on chip without feature map I/O to DRAM through layer fusion processing. The corresponding weight update can be directly loaded from internal weight SRAM through weight fusion to avoid DRAM latency.

\begin{table}
\centering
\caption{The architecture of our keyword spotting model}
\label{07_model}
\begin{tabular}{|l|l|} 
\hline
\rowcolor[rgb]{0.753,0.753,0.753} steps & input \\ 
\hline
preprocessing & high-pass filter, BN, quantize \\ 
\hline
convolution in CIM & (conv, max pooling) x5 \\ 
\hline
weight fusion & weight update \\ 
\hline
convolution in CIM & conv, max pooling, conv \\ 
\hline
post-processing & global average pooling \\
\hline
\end{tabular}
\end{table}

\section{Experimental Result} 
\subsection{Network Simulation Result} 
To showcase its efficacy, we adopt the keyword spotting task as an illustrative example.
The architecture of the keyword spotting model, as illustrated in Fig.~\ref{07_model}, comprises pre-processing, 1-D binary convolution, max pooling, and post-processing. The model achieves an accuracy of 94.02\% in the Google Speech Commands Dataset (GSCD)\cite{2018_GSCD} with 12 classes.

Simulated latency optimization has included DRAM access latency based on DDR4 DRAM~\cite{2016_Ramulator}. 
By means of layer fusion, the latency of the convolution execution can be effectively reduced by 33.16\% due to the decreased movement of the feature map between the DRAM and the chip.
Moreover, through weight fusion, an additional 62.94\% of latency can be saved by minimizing weight transfer between the DRAM and the chip. Subsequently, by employing a pipeline for convolution and max pooling operations, another 40.00\% of latency can be reduced by mitigating idle cycles of CIM macros.
In summary, the proposed optimizations result in a significant 85.14\% reduction in the end-to-end inference latency.

\subsection{Implementation Results and Comparison} 
The proposed digital design has been implemented with Verilog and synthesized with the Synopsys Design Compiler. Through the integration of the proposed design with the CIM macro, CIMR-V achieves 26.21 TOPS and 3707.84 TOPS/W with the TSMC 28nm CMOS process, when operating at a clock rate of 50 MHz.

Table~\ref{tb01_HW_comparison} presents a summary of performance metrics and a comparative analysis with other SRAM-based CIM designs.
In ~\cite{2021_JSSC_SRAM}, a dedicated dataflow is proposed for specific algorithms, which cannot be reconfigured for alternative algorithms.
Both ~\cite{2022_ISSCC_DIANA} and ~\cite{2022_TCAS_SRAM} did not consider weight fusion, resulting in increased latency in transferring weights between DRAM and chip.
Although ~\cite{2022_TCAS_SRAM} adopted multiple small CIM macros, which reduced energy efficiency due to overhead from peripheral circuits, the proposed design achieves higher energy efficiency and supports end-to-end model inference compared to other designs.

\section{Conclusion} 
This paper proposes an end-to-end CIM accelerator with RISC-V, which supports CIM-type instructions and full stack flow.
The entire design achieves an 85.14\% reduction in execution latency for the keyword spotting model through the use of CIM layer fusion, weight fusion, and pipeline with convolution and maximum pooling. 
The proposed design, which was implemented using TSMC 28nm CMOS technology, attains an impressive 3707.84 TOPS/W and 26.21 TOPS with a 50 MHz clock rate.

\bibliographystyle{IEEEtran}

\bibliography{bib/ieeeBSTcontrol,bib/thesis}

\end{document}